\documentclass[showpacs,preprintnumbers,amsmath,amssymb]{revtex4}
\usepackage{graphicx}
\usepackage{dcolumn}
\usepackage{bm}
\begin{document}

\title{Study of Gauge Symmetry of Some Field Theoretical Models Through the Lagrangian
Formulation }

\author{Safia Yasmin}
\affiliation{Indas Mahavidyalaya, Bankura - 722205, West Bengal,
India}
\author{Anisur Rahaman}
\email{anisur.rahman@saha.ac.in} \affiliation{Govt. College of
Engineering and Textile Technology, Serampore - 712201, Hooghly,
West Bengal, India}

\date{\today}

\begin{abstract}
Study of gauge symmetry is carried over the different interacting
and noninteracting field theoretical models through a prescription
based on lagrangian formulation. It is found that the prescription
is capable of testing whether a given model posses a gauge
symmetry or not. It can successfully formulate the gauge
transformation generator in all the cases whatever subtleties are
involved in it. It is found that the prescription has the ability
to show a direction how to extend the phase space using auxiliary
fields to restore the gauge invariance of a theory. Like the usual
phase space the prescription is found to be equally powerful in
the extended phase space of a theory.
\end{abstract}
 \maketitle
\section{Introduction}
Every basic interaction is supposed to have their origin from the
gauge principle
 and understanding of the gauge symmetry of a physical theory is a very important problem
  which has received much attention to the physicist from the long past. In a gauge theory,
   their exists some transformation that leaves physical content of the theory
   invariant. It even stands as a fundamental principle that determines the form of
    lagrangian of a theory. Two main approaches have been followed in the literature to study
     the local symmetry of the lagrangian of the gauge theories. The oldest one is the hamiltonian
     formulation based on
     Dirac conjecture \cite{sf0,sf1,sf2}. Several authors have tried to find out the answer of several
      interesting questions related to the gauge symmetry using hamiltonian formulation
      \cite{lag2,lag3,lag4, FLU,lag8,lag9,lag10,lag11,lag12,lag14}.
The most general form of gauge transformation generator too can be
determined with that hamiltonian formulation. To study BRST
symmetry, hamiltonian approach also has been found to be
instrumental \cite{lag15,lag16,lag17,lag18,lag19,lag21,
BRS,lag22}.

It is true that unitarity of a theory can not be well understood
without hamiltonian approach. However, hamiltonian embedding of
constrained system has some drawbacks. It does not always lead to
Lorentz covariant generating functional. This drawback indeed has
the remedy in the lagrangian formulation. So the importance of the
study of gauge symmetric property through the formalism based on
lagrangian formulation can not be ignored. Therefore, gauge
symmetry related studies on dynamical theory should be extended
with equal intensity in both the approaches.  Few studies using
lagrangian approach are available in the literature \cite{SUN,
GIT, lag23}. However, before Shirzad very little was achieved to
understand the fundamental question related to the gauge symmetry
in the lagrangian formulation. In \cite{lag23}, Shirzad  gave a
systematic development of gauge symmetry related study for an
arbitrary lagrangian and applied it to the so called generalized
Schwinger model \cite{MIO}. It is fair to admit that not much
attention have been paid in this direction after that. So
application of this formalism on the different field theoretical
models in order to test whether a given  model does have gauge
symmetry or it is lacking in it and hence to find out the
appropriate gauge transformation for that theory would be of
interest. It would be much more interesting to apply this
formalism in the extended phase space needed to restore the gauge
invariance and to verify whether this formalism works there in an
appropriate manner as it was found to work in the usual phase
space in \cite{lag23} would certainly be an interesting subject of
study.

The aim of this paper  is therefore, to study the different gauge
symmetric and gauge non symmetric (anomalous) models with the
prescription based on lagrangian formulation developed by Shirzad
in \cite{lag23}. We are intended to investigate whether Shirzad's
formalism enables one to verify the presence or absence of gauge
symmetry in a given theory. One reasons behind the consideration
of anomalous model is to ensure whether this scheme is capable of
testing the absence of gauge symmetry when it is lacking in a
given model. The another reason is to study the power of this
approach towards its applicability in the extended phase space.
Thus the application of this approach on anomalous models when
shows the absence of gauge symmetry, it is immediately modified in
such a way that the gauge symmetry gets restored and hence apply
the formalism to find out the correct gauge transformations in
that situation.

The paper is organized as follows. In  Sec. $II$, we have
presented briefly the general formalism to study the gauge
symmetry of a singular lagrangian developed by Shirzad in
\cite{lag23}. In Sec. $III$, we apply the above formulation on few
non-interacting models. The models which are not gauge invariant,
 are made gauge invariant by adding some appropriate terms
involving auxiliary fields and hence calculate the appropriate
gauge transformation generators. Sec. $IV$, is dealt with two
interacting field theoretical model e. g., Schwinger model with
mass like term for gauge field \cite{lag12,lag14} and Chiral
Schwinger model with Fadeevian\cite{lag10,lag11} anomaly in the
same perspective. The last section is containing a conclusion.
\section{A brief discussion of Shirzad's formalism}
 In order to make this paper self contained  we would like give a brief account of the
 formalism developed by Shirzad \cite{lag23} in this section. If a dynamical system
 with N degrees of freedom is considered
which is described by the lagrangian,
\begin{equation}
L= L(q_i, \dot{q}_i),
\end{equation}
the Euler equations of motion for that lagrangian will be
\begin{equation}
L_{i}=w^{}_{ij} \ddot{q_{j}} +\alpha_{i}.\label{lag}
\end{equation}
Here i=1,2.....N. The matrix $ w$ stands for the Hessian matrix of
the system. The Hessian matrix $w_{ij}$ and $\alpha_{i}$ of
equation (\ref{lag}) respectively are
\begin{equation}
w_{ij}=\frac{\partial^{2}L}{\partial\dot{q_i}\partial\dot{q_j}}.
\label{M1}
\end{equation}
\begin{equation}
\alpha_{i}=\frac{\partial^{2}L}{\partial{q_j}\partial\dot{q_i}}\dot{q_{j}}-\frac{\partial
L}{\partial{q_i}}.\label{M2}
\end{equation}
For a singular lagrangian $det[ w^{}_{ij}]=0$. The equations of
motion in this situation can not be solved for all accelerations.
If the rank of $ w$  is $N - A_{1}$ then $ A_{1}$ number of null
eigen vector will be found for the matrix $w^{}_{ij}$.
\begin{equation}
\lambda^{a_{1}}_{i} w_{ij}=0,
\end{equation}
where $ a_{1}=1...........A_{1}$. Here $\lambda^{a_{1}}_{i}$
indicates the null eigen vector. When equation (\ref{lag}) is
multiplied by the null eigen vector $\lambda^{a_{1}}_{i}$ from the
left it gives.
\begin{equation}
\gamma^{a_{1}}=\lambda^{a_{1}}_{i}L_{i}=\lambda^{a_{1}}_{i}\alpha_{i}=0.
\label{gama}
\end{equation}
This indicates the presence of $ A_{1}$ number of lagrangian
constraints of velocity and coordinates, but all of these
constraints are not independent of each other in general. If it is
assumed that the rank of equation (\ref{gama}) is $\bar A_{1}$,
then $\bar A_{1}$ number of independent functions
$\gamma^{\bar{a1}}$, can be written down as
\begin{equation}
\gamma^{\bar{a_{1}}}(q,
\dot{q})=\Sigma^{A_{1}}_{a_{1}=1}C^{\bar{a_{1}}}_{a_{1}} (q,
\dot{q}) \gamma^{a_{1}}(q,\dot{q}),    \label{bar}
\end{equation}
where ${\bar{a_{1}}}=1,......{\bar{A_{1}}}$, and
$C^{\bar{a_{1}}}_{a_{1}}$ represents the coefficients which may
depend on $q_{i}$ and $\dot{q_{i}}$. These set of lagrangian
constraints are useful for determining the number of undetermined
accelerations. Remaining constraints which vanish identically are
 \begin{equation}
\Sigma^{A_{1}}_{a_{1}=1}C^{\hat{a_{1}}}_{a_{1}} (q, \dot{q})
\gamma^{a_{1}}(q,\dot{q})=0,  \label{sum}
\end{equation}
where ${\hat{a_1}}=1,......{\hat{A_1}}.$
 These are of course linear combinations of $ \gamma^{a_{1}}$'s and their number
 will be $\hat{A_{1}}= A_{1}-\bar{A_{1}}$. This set of lagrangian constraints can be
  used to construct the form of the gauge transformation from lagrange equations of motion.
   Comparing equation $(\ref{gama})$ and (\ref{bar}) one gets,
 \begin{equation}
\lambda^{\bar{a_{1}}}(q,\dot
q)=\sum{^{A_{1}}_{a_{1}=1}}C^{\bar{a_{1}}}_{a_{1}} (q, \dot{q})
\lambda^{a_{1}}(q,\dot{q}).    \label{lam}
\end{equation}
Using equation (\ref{lam}), primary constraints can be calculated
and these are given by
\begin{equation}
\gamma^{\bar{a_{1}}}(q, \dot{q})=\lambda^{\bar{a_{1}}}_{i}L_{i}.
\label{CON}
\end{equation}
Equating equation (\ref{gama}) and (\ref{sum}) one can obtain
 \begin{equation}
\lambda^{\hat
a_{1}}=\sum{^{A_{1}}_{a_{1}=1}}C^{\hat{a_{1}}}_{a_{1}} (q,
\dot{q}) \lambda^{a_{1}}(q,\dot{q}). \label{EIGV}
\end{equation}
Equation (\ref{EIGV}), represents $\hat{A_{1}}$ number of null
eigen vector. Now identities of the Euler derivatives appears as
\begin{equation}
\lambda^{\hat a_{1}}_{i}L_{i}=0.
\end{equation}
In order to get a consistent theory, the time derivatives of the
primary constraint (\ref{CON}) is to be added to the equation of
motion (\ref{lag}). Therefore, one gets $ N+\bar{A_{1}}$ number of
equations that contain accelerations which can be written in a
combined manner as follows
\begin{equation}
L_{i_{1}}=w^{1}_{i_{1}j} \ddot{q_{j}} +\alpha^{1}_{i_{1}}=0,
\label{lag1}
\end{equation}
where $i_{1}=1..........N+\bar{A_{1}}$. Here $\bar{A_{1}}$
represents the rank of time derivatives of $\gamma^{\bar{a_{1}}}$.
The matrix $w^{1}_{i_{1}j}$ may also contain some other null eigen
vector like the previous one. Following the previous process one
gets new null eigen vector $ \lambda^{a_{2}}$, and the expressions
of it is
\begin{equation}
\gamma^{a_{2}}(q,\dot{q})=\lambda^{a_{2}}_{i_{1}}L^{1}_{i_{1}}=0,
\label{AB}
\end{equation}
where $a_{2}=1..........A_{2}$. So, one finds  $\bar{A_{2}}$
number of independent functions $\gamma^{\bar{a_{2}}}$ and
$\hat{A_{2}}$ number of identities $\gamma^{\hat{a_{2}}}$  for
$\gamma^{a_{2}}$ standing in equation (\ref{AB}). In the next step
the time derivative of secondary constraint is to be added to the
equation (\ref{lag1}), as it is done for the former set of
constraint in order to maintain consistency. This gives,
\begin{equation}
L_{i_{2}}=w^{2}_{i_{2}j} \ddot{q_{j}} +\alpha^{2}_{i_{2}}=0,
\label{im}
\end{equation}
where $i_{2}=1..........N+\bar{A_{1}}+\bar{A_{2}}$. Here
$\bar{A_{2}}$ stands for the rank of time derivative of secondary
constraint.
 In this way one needs to proceed step by step. In each step  some identities
 along with  some  new constraints may results. Finally,  in the $ n^{th}$ stage
the equations of motion for the system will be of the form
\begin{equation}
L_{n}=w^{n}_{i_{n}j}\ddot{q_{j}} +\alpha^{n}_{i_{n}}, \label{EQMN}
\end{equation}
where $i_{n}=1,..........,N+\bar{A_{1}}+\bar{A_{n}}$. If in this
case one finds new null eigen vector$ \lambda^{a_{n+1}} $ for $
w^{n}$  and multiplication of $ \lambda _{i_{n}}^{a_{n+1}} $ with
equation (\ref{EQMN}) provides $ \bar{A}_{n+1} $ number of
lagrangian constraints and  $\hat{A}_{n+1}$ number of identities
and these will hold the following relation.
\begin{equation}
\lambda^{\hat a_{n+1}}_{i} L_{i}+\lambda^{\hat
a_{n+1}}_{\bar{a_{1}}} \frac{d{\gamma^{\bar{a_{1}}}}}{dt}+.....
+\lambda^{\hat{a_{n+1}}}_{\bar{a_{n}}}\frac{d\gamma^{\bar{a_{n}}}}{dt}=0.
\label{lambda}
\end{equation}
Equation (\ref{lambda}) can be written down in the form of a total
derivative as follows
\begin{equation}
\sum^{n}_{s=0}\frac{d^{s}}{dt^{s}}(\phi_{si}L_{i})=0, \label{ph}
\end{equation}
where $\phi_{si}$ are some functions of coordinate and their
derivatives. That can be determined with a judicious choice. If $
w^{n}$ does not give any new eigen vector, it indicates that the
process gets terminated. Another way of testing the termination of
the procedure is to check whether the $ n^{th}$ step gives any new
constraint or not. The appearance of no new constraint too
indicate the termination of the process. For the lagrangian
$L(q_i, \dot{q}_i)$ the action is found to be invariant under the
following transformation,
\begin{equation}
\delta
q_{i}=\sum^{n}_{s=0}{(-1)}^{s}\frac{d^{s}f}{dt^{s}}\phi_{si}
\label{q},
\end{equation}
if $\phi_{si}$ exists for that particular dynamical system
represented by the $L(q_i, \dot{q}_i)$, where $f(t)$ is an
arbitrary function of time. The variation of lagrangian under the
transformation (\ref{q}) is given by
\begin{equation}
\delta {L}=-[\sum^{n}_{s=0}\frac{d^{s}}{dt^{s}}\phi_{si}L_{i}]f=0.
\end{equation}

If the Lagrangian of a dynamical system is described by the set of
fields $q_{i}(x,t)$ the general form of the lagrangian reads
\begin{equation}
L=\int dx L( q_{i}(x,t), \partial_{x} q_{i}(x,t), \partial_{t}
q_{i}(x,t).
\end{equation}
The equations of motion in this situation becomes
\begin{equation}
L_{i}(x,t)=\int dy  w_{ij}(x,y) \ddot{q_{j}}(y,t)+\alpha_{i}(x,t),
\end{equation}
where $ i=1,...............,N$.  N here represents the number of
fields describing the dynamical system. $w(x,y,t)$ and
$\alpha(x,t)$ in this situation takes the form
\begin{equation}
w_{ij}(x,y,t)=\frac{\delta^{2}L}{\delta \dot{q_{i}(x,t)}\delta
\dot{q_{j}}(y,t)},
\end{equation}
\begin{equation}
\alpha_{i}(x,t)=\int dy (\frac{\delta^{2}L}{\delta
q_{j}(y,t)\dot{\delta q_{j}}(x,t)}-\frac{\delta L}{\delta
q_{i}(x,t)}).
\end{equation}
The null eigen vector of the Hessian matrix $w(x,y,t)$ here looks
\begin{equation}
\lambda^{a}_{z}(x)=\lambda^{a}\delta(z-x).
\end{equation}
Multiplication of $\lambda^{a}_{z}(x)$ with equation of motion
gives primary lagrangian constraint as follows
\begin{equation}
\gamma^{a}=\int dx
\lambda_{i}^{a}\delta(z-x)L_{i}(x,t)=\lambda^{a}_{i}L_{i}(z,t).
\end{equation}
If the process continues in the similar manner as it is described
earlier keeping in mind that the system is described by field then
one will  arrive at the  following gauge transformation formula
\begin{equation}
\delta
q_{i}(x,t)=\sum^{m}_{\alpha=1}\sum_{s=0}^{n_{\alpha}}(-1)^{s} \int
dz \frac{\delta^{s}f_{\alpha}(z,t)}{\delta t^{s}}
\phi_{si}^{\alpha}(z,x).\label{GTF}
\end{equation}
\section{Non- interacting fields e. g., Free Maxwell lagrangian, Maxwell lagrangian with
mass like term for the gauge field and Free Chiral Boson}
 Let us consider the lagrangian density  of Free Maxwell field
\begin{equation}
{\cal L}_{FM}=-{\frac{1}{4}F_{\mu\nu}F^{\mu\nu}}.
\end{equation}
In $(1+1)$ dimension the lagrangian density reads
\begin{equation}
{\cal
L}_{FM}={\frac{1}{2}(\dot{A_{1}}^2+{A_{0}^{\prime}}^2-2A_{0}^{\prime}\dot{A_{1}}}).\label{LFM}
\end{equation}
The equations of motion for the field $ A_{0}$ and $ A_{1}$ that
come out from equation (\ref{lag}) for the lagrangian ${\cal
L}_{FM}$ are
\begin{equation}
L_{A_{0}}={A_{0}}^{\prime \prime}-\dot{A_{1}}^{\prime},
\end{equation}
\begin{equation}
L_{A_{1}}=\ddot {A_{1}}-\dot {A_{0}}^{\prime}.
\end{equation}
 For the lagrangian (\ref{LFM}) $ w$ and $\alpha$ respectively are
\begin{equation}
w=\left( \begin{array}{rr}
0 & 0 \\
0 & 1 \\
\end{array}  \right)\delta(z-x),
\end{equation}
\begin{equation}
\alpha=\left( \begin{array}{cc} {A_{0}}^{\prime
\prime}-\dot{A_{1}}^{\prime} \\ -\dot {A_{0}}^{\prime}
\end{array}  \right).
\end{equation}
 It is found that Hessian matrix $w$ has a null eigen vector
\begin{equation}
\lambda^{1}(z) =\left( \begin{array}{rr} 1,0
\end{array}  \right)\delta(z-x).
\end{equation}\\
Multiplying the equation (\ref{LFM}) from the left by $\lambda^{1}
$, we get the primary lagrangian constraint
\begin{equation}
\gamma^{1}(z,t)=\int dx
\delta(z-x)L_{A_{0}}(x,t)=L_{A_{0}}(z,t)=({A_{0}}^{\prime
\prime}-\dot{A_{1}}^{\prime})(z,t).
\end{equation}
Time derivatives of $\gamma^{1}$ yields
\begin{equation}
\frac{\partial}{\partial
t}\gamma^{1}(x,t)=(\dot{A_{0}}^{\prime\prime}-\ddot{A_{1}}^{\prime})(x,t).
\label{g}
\end{equation}
According to Shirzad's prescription  equation (\ref{g}) is to be
added with (\ref{LFM}) in order to maintain consistency condition
of the primary constraint and that results
\begin{equation}
L_{1}(x,t)=(L_{FM}+\frac{\partial}{\partial
t}\gamma^{1})(x,t).\label{L1}
\end{equation}
$ w^{1}$ and $\alpha^{1}$ as standing in equation (\ref{M1}) and
(\ref{M2}) are found out for this system as
\begin{equation}
w^{1}=\left( \begin{array}{rr}
0 & 0 \\
0 & 1 \\
0&-\partial/ \partial x
\end{array}  \right)\delta(z-x).
\end{equation}
\begin{equation}
\alpha^{1}=\left( \begin{array}{cc}
{A_{0}}^{\prime \prime}-\dot{A_{1}}^{\prime}\\
-\dot {A_{0}^{\prime}}\\
\dot {A_{0}^{\prime \prime}}\\
\end{array}  \right).
\end{equation}
A straight forward calculation shows  that  $w^{1}$ also has   a
null eigen vector
\begin{equation}
\lambda^{2}(z) =\left( \begin{array}{rrr} 0,\partial/ \partial x,1
\end{array}  \right)\delta(z-x).
\end{equation}
Multiplying the equation (\ref{L1}) from left by $\lambda^{2} $ we
find that  $\gamma^{2}$ comes out to be zero. Explicitly,
\begin{equation}
\gamma^{2}(z,t)=\int dx\lambda_{i_{2}}^{2}L_{i_{1}}(x,t)
=\lambda^{2}\alpha_{1}(z,t)=0.
\end{equation}
So, $\lambda^{2}$ does not give rise to any new  constraint. As a
result we can not increase the the rank of equation for
accelerations and $\gamma^{2}$ can be expressed in the following
form
\begin{equation}
\gamma^{2}(x,t)=(\frac{\partial}{\partial t}
L_{A_{0}}+\frac{\partial }{\partial x}L_{A_{1}})(x,t).
\label{maxf}
\end{equation}
 Comparing (\ref{maxf}) with equation (\ref{ph}) we get the non
 vanishing $\phi_{si}$'s:
\begin{equation}
\phi_{1,1}(z,x)=\delta(z-x),
\end{equation}
\begin{equation}
\phi_{0,2}(z,x)=\frac{\partial}{\partial z}\delta(z-x).
\end{equation}
The gauge transformation formula (\ref{GTF}) gives the following
gauge transformation for the field $A_{0}$ and $A_{1}$.
\begin{equation}
\delta A_{0}=-\int dz \frac{\partial}{\partial t}f(z,t)
\delta(z-x)=-\frac{\partial }{\partial t}f(x,t), \label{g1}
\end{equation}
\begin{equation}
\delta A_{1}=-\int dz(\frac{\partial}{\partial
z}\delta(z-x))f(z,t)=\frac{\partial }{\partial x}f(x,t).
\label{g2}
\end{equation}
A little algebra shows that the variation of ${\cal L}_{FM}$ is
\begin{eqnarray}
\delta {\cal L}_{FM}(x,t)=-\sum^{n}_{s=0}\frac{\partial^{s}}{\partial t^{s}}( \Phi_{si}L_{i})(x,t) \nonumber \\
=-[\frac{\partial}{\partial t} L_{A_{0}}+\frac{\partial} {\partial
x}L_{A_{1}}](x,t)\nonumber =0.
\end{eqnarray}
It is shows that the lagrangian (\ref{LFM}) is invariant under the
gauge transformation (\ref{g1})  and (\ref{g2}). It is the
expected result since it is known that the lagrangian (\ref{LFM})
is invariant under the transformation $A_{\mu} \rightarrow A_{\mu}
- \partial_{\mu} f$.
\subsection{Maxwell
lagrangian with mass like term} Let us now add the mass like term
$\frac{a^{2}}{2}A_{\mu}A^{\mu}$ with the Maxwell lagrangian and
apply the formalism to test whether it has the gauge symmetry or
not. So lagrangian with which we are going to start start our
analysis is
\begin{equation}
L_{MM}=\int({-\frac{1}{4}F_{\mu\nu}F^{\mu\nu}+\frac{a^{2}}{2}A_{\mu}A^{\mu}})dx.
\end{equation}
In $(1+1)$ dimension the lagrangian density takes the following
form
\begin{equation}
{\cal L}_{MM}={\frac{1}{2}(\dot{A_{1}}-
A_{0}^{\prime})^2+\frac{a^{2}}{2}(A_{0}^{2}-A_{1}^{2}}).
\label{free}
\end{equation}
The equations of motion for the field $ A_{0}$ and $ A_{1}$ that
come out from equation (\ref{lag}) for the lagrangian under
consideration are
\begin{equation}
L_{A_{0}}={A_{0}}^{\prime \prime}-\dot{A_{1}}^{\prime}-a^{2}A_{0},
\end{equation}
\begin{equation}
L_{A_{1}}=\ddot {A_{1}}-\dot {A_{0}}^{\prime}+a^{2}A_{1}.
\end{equation}
The matrices $ w$ and $\alpha$  for this modified lagrangian come
out to be
\begin{equation}
w=\left( \begin{array}{rr}
0 & 0 \\
0 & 1 \\
\end{array}  \right)\delta(z-x),
\end{equation}
\begin{equation}
\alpha=\left( \begin{array}{cc} {A_{0}}^{\prime
\prime}-\dot{A_{1}}^{\prime}-a^{2}A_{0} \\ -\dot
{A_{0}}^{\prime}+a^{2}A_{1}
\end{array}  \right).
\end{equation}
The Hessian matrix $w$ here gives the following null eigen vector
\begin{equation}
\lambda^{1}(z) =\left( \begin{array}{rr} 1,0
\end{array}  \right) \delta(z-x).
\end{equation}
Equation (\ref{free}) when multiplied from left by $\lambda^{1}$,
it results the  primary lagrangian constraint
\begin{equation}
\gamma^{1}(x,t)=({A_{0}}^{\prime
\prime}-\dot{A_{1}}^{\prime}-a^{2}A_{0})(x,t)=L_{A_{0}}(x,t).\label{PRC}
\end{equation}
We need the time derivatives of $\gamma^{1}$ to calculate
$L_{1}(x,t)$:
\begin{equation}
\frac{\partial}{\partial
t}\gamma^{1}(x,t)=(\dot{A_{0}}^{\prime\prime}-\ddot{A_{1}}^{\prime}
-a^{2}\dot A_{0})(x,t). \label{m}
\end{equation}
Now adding equation (\ref{m}) with $L_{MM}$ we get
\begin{equation}
L_{1}(x,t)=(L_{MM}+\frac{\partial}{\partial
t}\gamma^{1})(x,t).\label{lm}
\end{equation}
This $L_{1}(x,t)$ is to be used for further analysis in order to
maintain consistency of the primary constraint (\ref{PRC}). Using
equation (\ref{M1}) and (\ref{M2})  the matrices $w^{1}$ and
$\alpha^{1}$ for this system are calculated as follows.
\begin{equation}
w^{1}=\left( \begin{array}{rr}
0 & 0 \\
0 & 1 \\
0&-\partial/ \partial x
\end{array}  \right)\delta(z-x),
\end{equation}
\begin{equation}
\alpha^{1}=\left( \begin{array}{cc}
{A_{0}}^{\prime \prime}-\dot{A_{1}}^{\prime} -a^{2}\ A_{0} \\
-\dot {A_{0}^{\prime}} + a^{2}A_{1}\\
\dot {A_{0}^{\prime \prime}}-a^{2}\dot A_{0}\\
\end{array}  \right).
\end{equation}
A straight forward calculation shows  that $w^{1}$ has another
null eigen vector
\begin{equation}
\lambda^{2}(z) =\left( \begin{array}{rr} 0,\partial/ \partial x,1
\end{array}  \right)\delta(z-x).
\end{equation}
Multiplying the equation (\ref{lm}) from left by $\lambda^{2}$, we
get
\begin{equation}
\gamma^{2}(x,t)=\int dx
\lambda^{2}_{i_{2}}(x,t)=a^{2}(A_{1}^{\prime}-\dot A_{0})(x,t).
\end{equation}
which is non vanishing one. This $\lambda^{2}$ is nothing but the
 secondary constraint for this system. For maintaining
consistency we have added the time derivative of secondary
constraint to the equation (\ref{lm}) and obtain
\begin{equation}
L_{2}(x,t)=(L_{1}+\frac{\partial}{\partial t}\gamma^{2})(x,t).
\label{lm1}
\end{equation}
The matrix $ w^{2}$ and $\alpha^{2}$  standing in equation
(\ref{im}) will be the following for this particular situation.
\begin{equation}
w^{2}=\left( \begin{array}{rr}
0 & 0 \\
0 & 1 \\

0&-\partial/ \partial x\\
 -a^{2}&0
\end{array}  \right)\delta(z-x),
\end{equation}
\begin{equation}
\alpha^{2}=\left( \begin{array}{cc}
{A_{0}}^{\prime \prime}-\dot{A_{1}}^{\prime}-a^{2}A_{0} \\ -\dot {A_{0}}^{\prime}+a^{2}A_{1}\\
\dot {A_{0}^{\prime \prime}}-a^{2}\dot A_{0}\\
a^{2}\dot A_{1}^{\prime}
\end{array}  \right).
\end{equation}
We find that $w^{2}$ again gives a  null eigen vector
\begin{equation}
\lambda^{3}(z) =\left( \begin{array}{rr} 0,\partial/ \partial
x,1,0
\end{array}  \right)\delta(z-x).
\end{equation}
Multiplying the equation (\ref{lm1}) from left by $\lambda^{3}$,
we obtain
\begin{equation}
\gamma^{3}(x,t)=a^{2}(A_{1}^{\prime}-\dot
A_{0})(x,t)=\gamma^{2}(x,t).
\end{equation}
Note that $\gamma^{3}(x,t)$ and $\gamma^{2}(x,t)$ are identical.
So multiplication of $\lambda^{3}$ with $ L_{2}$ does not provide
any new constraint. Therefore, we can not increase the the rank of
equation for accelerations. Let us now try to write  $\gamma^{3}$
in the form of equation (\ref{lambda}).
\begin{equation}
\gamma^{3}(x,t)=(\frac{\partial}{\partial t} L_{A_{0}}+\frac
{\partial}{\partial x} L_{A_{1}})(x,t).\label{lm2}
\end{equation}
Equating equation (\ref{lm2}) with (\ref{ph}) we get $\phi_{si}$'s
and the non vanishing $\phi_{si}$'s are found out to be
\begin{equation}
\phi_{1,1}(z,x)=\delta(z-x),
\end{equation}
\begin{equation}
\phi_{0,2}(z,x)=\frac{\partial}{\partial z}\delta(z-x).
\end{equation}
The gauge transformation formula (\ref{GTF}) gives the following
gauge transformation for the field $A_{0}$ and $A_{1}$
\begin{equation}
\delta A_{0}=\int dz \frac{\partial}{\partial t} f(z,t)
\delta(z-x)=-\frac{\partial}{\partial t} f(x,t), \label{lm4}
\end{equation}
\begin{equation}
\delta A_{1}=-\int dz (\frac{\partial}{\partial z}\delta(z-x))
f(z,t)=\frac{\partial }{\partial x}f(x,t). \label{lm5}
\end{equation}
The variation of the lagrangian density (\ref{free}) under the
variation of the fields (\ref{lm4}) and (\ref{lm5})  is
\begin{eqnarray}
\delta{\cal L}(x,t)=&-&\sum^{n}_{s=0}\frac{d^{s}}{dt^{s}}( \Phi_{si}L_{i} )f (x,t) \nonumber  \\
 =&-&[\frac {d}{dt} L_{A_{0}}+\frac {d}{dx} L_{A_{1}}]f(x,t) \nonumber\\
=&-&a^{2}(A_{1}^{\prime}-\dot A_{0})f(x,t).\label{dre}
\end{eqnarray}
Since $\delta{\cal L}_{MM}$ does not vanish, there is no gauge
symmetry of the lagrangian density (\ref{free}). The result here
to does not go beyond our expectation since it is known that the
presence of mass like term breaks the gauge invariance of the free
Maxwell theory. In the following section we will proceed to study
the application of the Shirzad's formalism in the extended phase
space of this system.
\subsection{Maxwell Lagrangian with mass like term made Gauge
invariant} We have seen in the previous Section that the
lagrangian density (\ref{free}) is not invariant under the Gauge
transformation (\ref{lm4}) and (\ref{lm5}). So we add some terms
involving auxiliary fields $\theta$ with the  lagrangian
(\ref{free}) in order to make right hand side of the equation
(\ref{dre}) zero. Lagrangian density under consideration along
with with the appropriate terms needed to make equation
(\ref{dre}) zero is
\begin{equation}
{\cal L}_{EM}={\frac{1}{2}(\dot{A_{1}}-
A_{0}^{\prime})^2+\frac{ae^{2}}{2}(A_{0}^{2}-A_{1}^{2})+\frac{a}{2}(\dot{\theta^{2}}-\theta^{\prime^{2}})+
ae(\dot{\theta}A_{0}-\theta^{\prime} A_{1}}). \label{LAGM}
\end{equation}
Note that the term is nothing but the Wess-Zunino term which we
have needed to add to make equation (\ref{dre}) zero. The
equations of motion for the field $A_{0}$, $ A_{1}$ and $\theta$
are
\begin{equation}
L_{A_{0}}={A_{0}}^{\prime
\prime}-\dot{A_{1}}^{\prime}-ae^{2}A_{0}-ae \dot{\theta},
\end{equation}
\begin{equation}
L_{A_{1}}=\ddot {A_{1}}-\dot
{A_{0}}^{\prime}+ae^{2}A_{1}+ae\theta^{\prime},
\end{equation}
\begin{equation}
L_{\theta}=a\ddot{\theta}-a\theta^{\prime \prime}+ae\dot A_{0}-ae
A_{1}^{\prime}.
\end{equation}
Here we repeat the same calculation as before and $w$ and $\alpha$
for this lagrangian are found out to be
\begin{equation}
w=\left( \begin{array}{rrr}
0 & 0& 0 \\
0 &1& 0 \\
0& 0& a\\
\end{array}  \right)\delta(z-x),
\end{equation}
\begin{equation}
\alpha=\left( \begin{array}{cc}
{A_{0}}^{\prime \prime}-\dot{A_{1}}^{\prime}-ae^{2}A_{0}-ae\dot{\theta} \\ -\dot {A_{0}}^{\prime}+ae^{2}A_{1}+ae\theta^{\prime}\\
-a\theta^{\prime \prime}+ae \dot A_{0}-aeA_{1}^{\prime}
\end{array}  \right).
\end{equation}
We also find that the Hessian matrix $w$ has the null eigen vector
\begin{equation}
\lambda^{1}(z) =\left( \begin{array}{rrr} 1,0 ,0
\end{array}  \right) \delta(z-x).
\end{equation}
Like the massless situation  we calculate the primary lagrangian
constraint in this situation too:
\begin{equation}
\gamma^{1}(x,t)=({A_{0}}^{\prime
\prime}-\dot{A_{1}}^{\prime}-ae^{2}A_{0}-ae\dot{\theta})(x,t).
\end{equation}
$ L_{1}$ in this situation is obtained as
\begin{equation}
L_{1}(x,t)=(L_{EM}+\frac{\partial}{\partial t}\gamma^{1})(x,t).
\label{eta4}
\end{equation}
Here $ w^{1}$ and $\alpha^{1}$  too  are found out using equation
(\ref{M1}) and (\ref{M2}).
\begin{equation}
w^{1}=\left( \begin{array}{rrr}
0 & 0 & 0\\
0 & 1 &0\\
0&0& a\\
0&-\partial/ \partial x&-ae
\end{array}  \right)\delta(z-x),
\end{equation}
\begin{equation}
\alpha^{1}=\left( \begin{array}{cc}
{A_{0}}^{\prime \prime}-\dot{A_{1}}^{\prime}-ae^{2}A_{0}-ae\dot{\theta} \\ -\dot {A_{0}}^{\prime}+ae^{2}A_{1}+ae\theta^{\prime}\\
-a\theta^{\prime \prime}+ae \dot A_{0}-aeA_{1}^{\prime}\\
\dot A_{0}^{\prime\prime}-ae^{2}\dot A_{0}
\end{array}  \right).
\end{equation}
 A little algebra shows that $w^{1}$ has the following null eigen vector
\begin{equation}
\lambda^{2}(z) =\left( \begin{array}{rr} 0,\partial/ \partial
x,+e,1
\end{array}  \right)\delta(z-x).
\end{equation}
Multiplying the equation (\ref{eta4}) from left by $\lambda^{2}$,
we find that
\begin{equation}
\gamma^{2}(x,t)=0.
\end{equation}
So, $\lambda^{2}$ does not give rise to any new  constraint.
Therefore, the process gets terminated. Thus the increase of the
rank of equation for acceleration is not possible here.
$\gamma^{2}$ here also can be written in the form of equation
(\ref{lambda}) as follows
\begin{equation}
\gamma^{2}(x,t)=(\frac {\partial}{\partial t} L_{A_{0}}+\frac
{\partial}{\partial x} L_{A_{1}}+eL_{\theta})(x,t). \label{eta5}
\end{equation}
Comparing equation (\ref{eta5}) with (\ref{ph}) we get the non
vanishing components of $\phi_{si}$'s:
\begin{equation}
\phi_{1,1}(z,x)=\delta(z-x),
\end{equation}
\begin{equation}
\phi_{0,2}(z,x)=\frac{\partial}{\partial z}\delta(z-x),
\end{equation}
\begin{equation}
\phi_{0,3}(z,x)=+e\delta(z-x).
\end{equation}
Finally we  find the gauge transformation of the fields for the
system with the help of equation (\ref{GTF}).
\begin{equation}
\delta A_{0}=-\frac{\partial }{\partial t} f(x,t), \label{eta12}
\end{equation}
\begin{equation}
\delta A_{1} =-\frac{\partial}{\partial x} f (x,t),\label{eta13}
\end{equation}
\begin{equation}
\delta {\theta}=+ef(x,t). \label{eta14}
\end{equation}
The variation of the lagrangian density (\ref{LAGM}) under the
transformation (\ref{eta12}), (\ref{eta13}) and (\ref{eta14})
comes out to be
\begin{eqnarray}
\delta {\cal L}_{EM} &=&-\sum^{n}_{s=0}\frac{d^{s}}{dt^{s}}( \Phi_{si}L_{i} )f(x,t)\nonumber \\
&=&-[\frac{\partial}{\partial t}L_{A_{0}}+\frac{\partial}{\partial
x}L_{A_{1}}+eL_{\theta}]f(x,t) =0. \label{AC}
\end{eqnarray}
Equation ({\ref{AC})  confirms that the action is invariant under
the gauge transformation (\ref{eta12}), (\ref{eta13}) and
(\ref{eta14}).  Note that the formalism shows its successful
application in the extended phase space of this simple non
interacting system. In the following section the formalism is
again applied to another  noninteracting field theory e.g., Chiral
boson which is known as a basic ingredient of heterotic string
theory .
\subsection{Free Chiral Boson}
Free Chiral Boson\cite{FLU,sig,bell,PPS} though a very simple
field theory  the study of Gauge symmetry for this system is very
subtle and interesting because the lagrangian of Chiral Boson
contains a second class constraint
$(\partial_{0}+\partial_{1})\phi=0$. So it is studied here using
Shirzad's formalism. Lagrangian density of free Chiral Boson  as
described in \cite{PPS, PPS1} is given by
\begin{equation}
{\cal
L}_{CB}=\frac{1}{2}(\dot{\phi^{2}}-\phi^{\prime^{2}})+\eta(\dot{\phi}-\phi^{\prime})\label{sa}
\end{equation}
Here $\eta$ stands for the lagrange multiplier field. The
equations of motion for the field $\phi$ and $ \eta$ are
\begin{equation}
L_{\phi}=(\ddot {\phi}-\phi^{\prime \prime})
+\dot{\eta}-\eta^{\prime}
\end{equation}
\begin{equation}
L_{\eta}=-(\dot{\phi}-\phi^{\prime})
\end{equation}
In this case $ w$ and $\alpha$ are,
\begin{equation}
w=\left( \begin{array}{rr}
1& 0  \\
0 & 0 \\
\end{array}  \right)\delta(z-x),
\end{equation}
\begin{equation}
\alpha=\left( \begin{array}{cc}
-\phi^{\prime\prime}+(\dot{\eta}-\eta^{\prime})\\
-\dot{\phi}+\phi^{\prime}
\end{array}  \right).
\end{equation}
A little algebra shows that Hessian matrix $w$ has the null eigen
vector, $\lambda^{1}(z) =\left( \begin{array}{rr} 0,1
\end{array}  \right) \delta(z-x)$.
Multiplying $\lambda^{1}$ with equation (\ref{sa})  from left we
obtain the primary constraint of the theory as usual.
\begin{equation}
\gamma^{1}(x,t)=(-\dot{\phi}+\phi^{\prime})(x,t).
\end{equation}
The consistency of this primary constraint with time needs to be
maintained which necessities to calculate  $ L_{1}$  as follows
for further analysis
\begin{equation}
L_{1}(x.t)=(L_{CB}+\frac{\partial \gamma^{1}}{\partial t})(x,t).
\label{wesb}
\end{equation}
After a little algebra, we have $w^{1}$ and $\alpha^{1}$:
\begin{equation}
w^{1}=\left( \begin{array}{rr}
1 & 0 \\
0 & 0 \\
-1& 0\\

\end{array}  \right)\delta(z-x),
\end{equation}
\begin{equation}
\alpha^{1}=\left( \begin{array}{cc}
-\phi^{\prime\prime}+(\dot {\eta}-\eta^{\prime})\\
-\dot{\phi}+{\phi}^{\prime}\\
\dot{\phi}^{\prime}
\end{array}  \right).
\end{equation}
We find that $w^{1}$ has another null eigen vector $\lambda^{2}
=\left(
\begin{array}{rr} 1,0,1
\end{array}  \right)\delta(z-x)$.
So secondary lagrangian constraint is now obtained, multiplying
$\lambda^{2}$ with (\ref{wesb}).
\begin{equation}
\gamma^{2}(x,t)=(-\phi^{\prime\prime}+\dot{\eta}-\eta^{\prime}+\dot\phi^{\prime})(x,t)
\end{equation}
Adding the time derivatives of $\gamma^{2}$  with $ L_{1}$,
$L_{2}$ is obtained to maintain the consistency of the secondary
constraint with time.
\begin{equation}
L_{2}(x,t)=(L_{1}+\frac{\partial\gamma^{2} }{\partial t})(x,t).
\label{boso6}
\end{equation}
For $L_{2}$, the matrices $w^{2}$ and $\alpha^{2}$ are found out
as
\begin{equation}
w^{2}=\left( \begin{array}{rrr}
1 & 0\\
0 & 0\\
-1&0\\
\partial/ \partial x&1

\end{array}  \right)\delta(z-x),
\end{equation}
\begin{equation}
\alpha^{2}=\left( \begin{array}{cc}

-\phi^{\prime\prime}+(\dot {\eta}-\eta^{\prime})\\
-\dot{\phi}+{\phi}^{\prime}\\
+\dot{\phi}^{\prime}\\
-\dot{\phi}^{\prime\prime}-\dot{\eta}^{\prime}
\end{array}\right).
\end{equation}
It is found that $w^{2}$ is also having a null eigen vector
$\lambda^{3}(z) =\left(
\begin{array}{rr} 1,0,1,0
\end{array}  \right)\delta(z-x)$.
Multiplying the equation (\ref{boso6}) from left by $\lambda^{3}$
we find
\begin{equation}
\gamma^{3}(x,t)=\gamma^{2}(x,t). \label{boso8}
\end{equation}
So from the previous step it can be concluded that there is no
further constraint and we can not increase the rank of equations
for accelerations.  The process is thus terminated. We now  write
$\gamma^{3}$ in the following form.
\begin{equation}
\gamma^{3}(x,t)=(L_{\phi}+\frac{\partial}{\partial
t}L_{\eta})(x,t) \label{nm}
\end{equation}
At this stage we need to compare  (\ref{nm}) with equation
(\ref{ph}) to compute the following non vanishing $\phi_{si}$'s.
\begin{equation}
\phi_{0,1}(z,x)=\delta(z-x),
\end{equation}
\begin{equation}
\phi_{1,2}(z,x)=\delta(z-x).
\end{equation}
Finally we obtain the gauge transformation of the field $\phi$ and
$\eta$ for the system using equation (\ref{GTF}).
\begin{equation}
\delta \phi=f(x,t),\label{GA}
\end{equation}
\begin{equation}
\delta\eta=-\frac{\partial}{\partial t}f(x,t).\label{GB}
\end{equation}
Let us now calculate the variation of ${\cal L}$ under the above
transformations of the fields (\ref{GA}) and (\ref{GB}).
\begin{equation}
\delta {\cal L}_{CB}=-(L_{\phi}+\frac{\partial L_{\eta}}{\partial
t})(x,t)=-(-\phi^{\prime\prime}+\dot{\eta}-\eta^{\prime}+\dot{\phi}^{\prime})f(x,t).
\label{simi}
\end{equation}
This shows that the lagrangian (\ref{sa}) is not invariant under
the above gauge transformations. The result of course have not
gone beyond our expectation because chiral boson is known not to
possess any gauge symmetry. This shows that the formalism is
capable of testing the gauge symmetric property of this simple
system having subtlety in many respects.
\subsection{Free Chiral Boson in the Extended phase space} Let us
add some appropriate terms involving auxiliary fields $\theta$ to
the lagrangian density of free Chiral Boson that makes the right
hand side of the equation (\ref{simi}) zero. It is found that
Lagrangian density that satisfy the above requirement is
\begin{equation}
{\cal
L}_{ECB}=\frac{1}{2}(\dot{\phi^{2}}-\phi^{\prime^{2}})+\eta(\dot{\phi}-\phi^{\prime})
-\frac{1}{2}(\dot{\theta}^{2}+\theta^{\prime^{2}})+\phi^{\prime}\theta^{\prime}
+\dot{\theta}\theta^{\prime}-\dot{\theta}\phi^{\prime}-\eta(\dot{\theta}-\theta^{\prime}).\label{bow12}
\end{equation}
What follows next is to study the  gauge symmetric property of the
lagrangian (\ref{bow12}) using the formalism given in section
$(2)$. To this end we calculate the equations of motion
corresponding to the field $\phi$, $ \eta$ and $\theta$
\begin{equation}
L_{\phi}=(\ddot {\phi}-\phi^{\prime \prime})
+\dot{\eta}-\eta^{\prime}+\theta^{\prime\prime}-\dot{\theta}^{\prime},
\end{equation}
\begin{equation}
L_{\eta}=-(\dot{\phi}-\phi^{\prime})+\dot{\theta}-\theta^{\prime},
\end{equation}
\begin{equation}
L_{\theta}=-\ddot{\theta}-\theta^{\prime\prime}+{\phi^{\prime\prime}}+2\dot{\theta}^{\prime}
-\dot{\phi^{\prime}}-\dot{\eta}+\eta^{\prime}.
\end{equation}
 In this situation $ w$ and $\alpha$ are
\begin{equation}
w=\left( \begin{array}{rrr}
1& 0&0  \\
0 & 0&0\\
0&0&-1\\
\end{array}  \right)\delta(z-x),
\end{equation}
\begin{equation}
\alpha=\left( \begin{array}{ccc}
-\phi^{\prime\prime}+(\dot{\eta}-\eta^{\prime})+\theta^{\prime\prime}-\dot{\theta}^{\prime}\\
-\dot{\phi}+\phi^{\prime}+\dot{\theta}-\theta^{\prime}\\

-\theta^{\prime\prime}+2\dot{\theta}^{\prime}+\phi^{\prime\prime}-\dot{\phi}^{\prime}+\eta^{\prime}-\dot{\eta}\\
\end{array}  \right).
\end{equation}
We find that $w$ has the following null eigen vector
\begin{equation}
\lambda^{1}(z) =\left( \begin{array}{rr} 0,1,0
\end{array}  \right) \delta(z-x).
\end{equation}
Multiplying $\lambda^{1}$ with equation (\ref{bow12}) from left,
we obtain the primary constraint in this situation
\begin{equation}
\gamma^{1}(x,t)=(-\dot{\phi}+\phi^{\prime}+\dot{\theta}-\theta^{\prime})(x,t).
\end{equation}
It is needed to add the time derivatives of $\gamma^{1}$ with $L$
for further analysis otherwise we will fail to maintain
consistency of the primary constraint.
\begin{equation}
L_{1}(x,t)=(L_{ECB}+\frac{\partial}{\partial t}\gamma^{1})(x,t).
\label{boso3}
\end{equation}
 $w^{1}$ and $\alpha^{1}$  here are
\begin{equation}
w^{1}=\left( \begin{array}{rrr}
1 & 0&0 \\
0 & 0&0 \\
0& 0&-1\\
-1&0&1\\
\end{array}  \right)\delta(z-x),
\end{equation}
\begin{equation}
\alpha^{1}=\left( \begin{array}{cc}
-\phi^{\prime\prime}+(\dot{\eta}-\eta^{\prime})+\theta^{\prime\prime}-\dot{\theta}^{\prime}\\
-\dot{\phi}+\phi^{\prime}+\dot{\theta}-\theta^{\prime}\\

-\theta^{\prime\prime}+2\dot{\theta}^{\prime}-\dot{\eta}+\eta^{\prime}+\phi^{\prime\prime}-\dot{\phi^{\prime}}\\
\dot{\phi}^{\prime}-\dot{\theta}^{\prime}
\end{array}  \right).
\end{equation}
A  little algebra shows that $w^{1}$ has a new null eigen vector
\begin{equation}
 \lambda^{2}(z) =\left( \begin{array}{rrrr}
1,0,1,1
\end{array}  \right)\delta(z-x).
\end{equation}
Multiplying $\lambda^{2}$ with equation (\ref{boso3}) we get the
following vanishing condition:
\begin{equation}
\gamma^{2}(x,t)=0.
\end{equation}
If we proceed to write $\gamma^{2}$  in the form of equation
(\ref{lambda}) we reach at
\begin {equation}
\gamma^{2}(x,t)=(L_{\phi}+\frac{\partial}{\partial
t}L_{\eta}+L_{\theta})(x,t).\label{er}
\end{equation}
Comparing the above equation (\ref{er}) with equation (\ref{ph})
we get the non vanishing $\phi_{si}$'s.
\begin{equation}
\phi_{0,1}(z,x)=\delta(z-x),
\end{equation}
\begin{equation}
\phi_{1,2}(z,x)=\frac{\partial}{\partial t} \delta(z-x),
\end{equation}
\begin{equation}
\phi_{0,3}(z,x)=\delta(z-x).
\end{equation}
We  are now in a position to compute the gauge transformation for
the fields describing the system:
\begin{equation}
\delta \phi=f(x,t), \label{sz}
\end{equation}
\begin{equation}
\delta\eta=-\frac{\partial}{\partial t} f(x,t), \label{sz1}
\end{equation}
\begin{equation}
\delta \theta=-f(x,t). \label{sz2}
\end{equation}
A little algebra shows that ${\cal L}_{ECB}$ will show the
following variation under the above set of transformations
(\ref{sz}), (\ref{sz1}) and (\ref{sz2}).
\begin{equation}
\delta {\cal L}_{ECB}=-(L_{\phi}+\frac{\partial}{\partial
t}L_{\eta}+L_{\theta})f(x,t)=0.
\end{equation}
It shows that the lagrangian (\ref{bow12}) is invariant under the
transformation (\ref{sz}), (\ref{sz1}) and (\ref{sz2}). It is the
expected result because the terms which we are forced to add to
make equation (\ref{simi}) zero is nothing but the Wess-Zumino
term that has brought back the gauge symmetry in the system. Thus
the formalism is found to work successfully in the extended phase
space of this system too. So for free field theories the formalism
is found to works equally well both in the usual and extended
phase space. In the following sections we will consider some
interacting system to test how well it woks there .
\section{Vector Schwinger model with  mass like term for the gauge field}
Vector Schwinger model \cite{lag26,lag27,lag28} is an interesting
field theoretical model which though posses gauge symmetry
however, the same model with mass like term for the Gauge field as
studied in \cite{lag12,lag14,BRS} does not posses that symmetry.
It is an interesting model where Gauge boson acquires mass in the
same way as it acquires in the Chiral Schwinger model \cite{lag2,
lag3,lag4} and the fermion gets liberated. So it would be of worth
to test the gauge symmetric property of this model through this
prescription. We would like to mention here that the usual vector
Schwinger model was tested by Shirzad through his prescription as
a limiting case \cite{lag23} of the generalized Schwinger model
\cite{MIO}. The lagrangian density in our present consideration is
\begin{equation}
{\cal L}_{SM}
=-\frac{1}{4}F_{\mu\nu}F^{\mu\nu}+\frac{1}{2}\partial_{\mu}\phi
\partial^{\mu}\phi+\frac{1}{2}ae^{2}A_{\mu}
A^{\mu}+\epsilon^{\mu\nu}A_{\mu}\partial_{\nu}\phi. \label{ceta}
\end{equation}
When it is written explicitly the lagrangian density reads
\begin{equation}
{\cal L}_{SM}=\frac{1}{2}(\dot{A_{1}}-
A_{0}^{\prime})^{2}+\frac{ae^{2}}{2}(A_{0}^{2}-A_{1}^{2})+\frac{1}{2}(\dot{\phi^{2}}-\phi^{\prime^{2}})+
(\phi^{\prime}A_{0}-\dot{\phi} A_{1}). \label{ew}
\end{equation}
The equations of motion for the field $\phi$ , $A_{0}$ and $
A_{1}$ for the lagrangian density under consideration are
\begin{equation}
L_{\phi}=(\ddot {\phi}-\phi^{\prime \prime})
+A_{0}^{\prime}-\dot{A_{1}},
\end{equation}
\begin{equation}
L_{A_{0}}={A_{0}}^{\prime \prime}-\dot{A_{1}}^{\prime}-
{\phi}^{\prime}-ae^{2}A_{0},
\end{equation}
\begin{equation}
L_{A_{1}}=\ddot {A_{1}}-\dot
{A_{0}}^{\prime}+ae^{2}A_{1}+\dot{\phi}.
\end{equation}
For this system the matrices $w$ and $\alpha$ are
\begin{equation}
w=\left( \begin{array}{rrr}
1& 0 & 0 \\
0 & 0 & 0 \\
0 & 0 & 1\\
\end{array}  \right)\delta(z-x),
\end{equation}
\begin{equation}
\alpha=\left( \begin{array}{cc}
-\phi^{\prime\prime}-\dot {A_{1} }+A_{0}^{\prime}\\
{A_{0}}^{\prime \prime}-\dot{A_{1}}^{\prime}- {\phi}^{\prime} -ae^{2}A_{0} \\ -\dot {A_{0}}^{\prime}+ae^{2}A_{1}+\dot{\phi}\\
\end{array}  \right).
\end{equation}
It is found that Hessian matrix $ w$ has a null eigen vector which is given by
\begin{equation}
\lambda^{1}(z) =\left( \begin{array}{rr} 0,1,0
\end{array}  \right)\delta(z-x),
\end{equation}
and we get the primary lagrangian constraint multiplying equation
(\ref{ceta}) by $\lambda^{1}$.
\begin{equation}
\gamma^{1}(x,t)=({A_{0}}^{\prime
\prime}-\dot{A_{1}}^{\prime}-ae^{2}A_{0}- {\phi}^{\prime})(x,t).
\end{equation}
To maintain consistency of the primary constraint with  time the time derivative of $\gamma^{1}$ is added
 with ${\cal L}_{SM}$ that gives
  $L_{1}$, which contains $w_{1}$ and $\alpha_{1}$ and those are
\begin{equation}
L_{1}(x,t)=(L_{SM}+\frac{\partial\gamma^{1}}{\partial t})(x,t),
\label{ceta3}
\end{equation}
\begin{equation}
w^{1}=\left( \begin{array}{rrr}
1 & 0 &0\\
0 & 0 & 0 \\
0& 0&1\\
0& 0&-\partial/ \partial x
\end{array}  \right)\delta(z-x),
\end{equation}
\begin{equation}
\alpha^{1}=\left( \begin{array}{cc}
-\phi^{\prime\prime}-\dot {A_{1} }+A_{0}^{\prime}\\
{A_{0}}^{\prime \prime}-\dot{A_{1}}^{\prime}- {\phi}^{\prime} -ae^{2}A_{0} \\
-\dot {A_{0}}^{\prime}+ae^{2}A_{1}+\dot{\phi}\\
\dot {A_{0}}^{\prime\prime}-ae^{2}\dot {A_{0}}-\dot\phi^{\prime}\\
\end{array}  \right).
\end{equation}
The matrix  $w^{1}$ is found to have a different null eigen vector
\begin{equation}
\lambda^{2}(z) =\left( \begin{array}{rr} 0,0,\partial/ \partial
x,1
\end{array}  \right)\delta(z-x).
\end{equation}
Multiplying equation( \ref{ceta3}) from left by $\lambda^{2}$ we
get the secondary constraint
\begin{equation}
\gamma^{2}(x,t)=ae^{2}(A_{1}^{\prime}-\dot{A_{0}})(x,t).
\end{equation}
The process is again repeated since in this situation it does not
lead to any terminating condition. Addition of the time derivative
of $\gamma^{2}$ with $ L_{1}$, we get $L_{2}$ as usual.
\begin{equation}
L_{2}(x,t)=(L_{1}+\frac{\partial}{\partial t}\gamma^{2})(x,t).
\label{ceta11}
\end{equation}
It contains $ w^{2}$ and $\alpha^{2}$ those are found out as
\begin{equation}
w^{2}=\left( \begin{array}{rrr}
1 & 0 &0\\
0 & 0 & 0 \\
0& 0&1\\
0&0&-{\partial/\partial x}\\
0&-ae^{2}&0
\end{array}  \right)\delta(z-x),
\end{equation}
\begin{equation}
\alpha^{2}=\left( \begin{array}{cc}
-\phi^{\prime\prime}-\dot {A_{1} }+A_{0}^{\prime}\\
{A_{0}}^{\prime \prime}-\dot{A_{1}}^{\prime}- {\phi}^{\prime} -ae^{2}A_{0} \\
-\dot {A_{0}}^{\prime}+ae^{2}A_{1}+\dot{\phi}\\
\dot {A_{0}}^{\prime\prime}-ae^{2}\dot {A_{0}}-\dot\phi^{\prime}\\
ae^{2}\dot{A_{1}}^{\prime}
\end{array}  \right).
\end{equation}
The matrix $w^{2}$ also is found to give another null eigen vector
\begin{equation}
\lambda^{3}(z) =\left( \begin{array}{rr} 0,0,\partial/ \partial
x,1,0
\end{array}  \right) \delta(z-x).
\end{equation}
On multiplying  equation (\ref{ceta11}) from left by $\lambda^{3}$
we find
\begin{equation}
\gamma^{3}(x,t)=ae^{2}(A_{1}^{\prime}-\dot{A_{0}})(x,t)
=\gamma^{2}(x,t).\label{TC}
\end{equation}
Equation (\ref{TC}) indicates a terminating condition. So
multiplication of $\lambda^{3} $ with $ L_{2} $ does not gives
rise to any new  constraint. The rank of equation for
accelerations therefore, does not increase here. One can express
$\gamma^{3}$ in the form of equation (\ref{lambda}) in a
straightforward manner.
\begin{equation}
\gamma^{3}(x,t)=(\frac {\partial}{\partial t} L_{A_{0}}+\frac
{\partial}{\partial x} L_{A_{1}})(x,t).\label{ziz}
\end{equation}
In order to get the non vanishing $\phi_{si}$'s we equate equation
(\ref{ziz}) with  (\ref{ph}) and
 we find that
\begin{equation}
\phi_{0,1}(x,z)=0,
\end{equation}
\begin{equation}
\phi_{1,2}(z,x)=\delta(z-x),
\end{equation}
\begin{equation}
\phi_{0,3}(z,x)=\frac{\partial}{\partial x}\delta(z-x).
\end{equation}
The gauge transformation formula (\ref{GTF}) here renders the
following gauge transformation for the field $\phi$, $A_{0}$ and
$A_{1}$:
\begin{equation}
\delta \phi=0, \label{wt}
\end{equation}
\begin{equation}
\delta A_{0}=-\frac{\partial }{\partial t}f(x,t),\label{wt1}
\end{equation}
\begin{equation}
\delta A_{1}=-\frac{\partial }{\partial x}f(x,t).\label{wt2}
\end{equation}
The variation of ${\cal L }_{SM}$ under the transformations
(\ref{wt}), (\ref{wt1}) and (\ref{wt2}) is
\begin{eqnarray}
\delta{\cal L}_{SM}&=-&\sum^{n}_{s=0}\frac{\partial^{s}}{\partial t^{s}}( \Phi_{si}L_{i})f(x,t)\nonumber\\
&=&-(\frac {\partial}{\partial t} L_{A_{0}}+\frac {\partial}{\partial x} L_{A_{1}})f(x,t) \nonumber\\
&=&-ae^{2}(A_{1}^{\prime}-\dot{A_{0}})f(x,t).
\end{eqnarray}
The lagrangian density (\ref{ew}) is therefore, not invariant
under the transformation (\ref{wt}), (\ref{wt1}) and (\ref{wt2}).
Here also it appears that the formalism has tested correctly that
the lagrangian (\ref{ew}) has no gauge invariance in the usual
phase space.
\subsection{Vector Schwinger model with mass like term made Gauge invariant in the extended phase space}
In the preceding section we have found that the Shirzad's
prescription have successfully tested that the lagrangian of the
Schwinger model with mass like term has no gauge invariance under
the transformation generated there. Let us add the appropriate
term to the lagrangian (\ref{ceta}) which certainly extend the
phase space of the theory and  investigate whether the process can
infer that gauge invariance has restored in the extended phase
space. The lagrangian under our present  consideration is
\begin{eqnarray}
L_{ESM}&=&\int dx[
-\frac{1}{4}F_{\mu\nu}F^{\mu\nu}+\frac{1}{2}\partial_{\mu}\phi
\partial^{\mu}\phi+\frac{1}{2}ae^{2}A_{\mu}
A^{\mu}\nonumber\\&+&\epsilon^{\mu\nu}\partial_{\mu}\phi
A_{\nu}-\frac{a}{2}\partial_{\mu}\theta
\partial^{\mu}\theta+aeA_{\mu}\partial^{\mu}\theta]. \label{bu}
\end{eqnarray}
Lagrangian density takes the following form in $(1+1)$ dimension
\begin{eqnarray}
{\cal L}_{ESM}= &+&\frac{1}{2}(\dot{A_{1}}-
A_{0}^{\prime})^{2}+\frac{ae^{2}}{2}(A_{0}^{2}-A_{1}^{2})+\frac{1}{2}(\dot{\phi^{2}}
-\phi^{\prime^{2}})\nonumber\\&+& (\phi^{\prime}A_{0}-\dot{\phi}
A_{1})+\frac{a}{2}(\dot{\theta}^{2}-
\theta^{\prime^{2}})+ae(A_{0}\dot{\theta}-A_{1}\theta^{\prime})
\label{qa}
\end{eqnarray}
The field $\theta$ represents an auxiliary field. The equations of
motion for the field $\phi$, $ A_{0}$, $ A_{1}$ and $\theta$ that
come out using equation (\ref{lag}) are
\begin{equation}
L_{\phi}=\ddot{\phi}-\phi^{\prime
\prime}+A_{0}^{\prime}-\dot{A_{1}},
\end{equation}
\begin{equation}
L_{A_{0}}={A_{0}}^{\prime \prime}-\dot{A_{1}}^{\prime}-
{\phi}^{\prime}-ae^{2}A_{0}-ae\dot{\theta},
\end{equation}
\begin{equation}
L_{A_{1}}=\ddot {A_{1}}-\dot
{A_{0}}^{\prime}+ae^{2}A_{1}+\dot{\phi}+ae\theta^{\prime},
\end{equation}
\begin{equation}
L_{\theta}=a(\ddot {\theta}-\theta^{\prime \prime})
+ae\dot{A_{0}}-aeA_{1}^{\prime}.
\end{equation}
For the lagrangian (\ref{qa}), $w$ and $\alpha$  are found out as
\begin{equation}
w =\left( \begin{array}{rrrr}
1& 0 & 0&0 \\
0 & 0 & 0 &0\\
0 & 0 & 1&0\\
0&0&0&+a\\
\end{array}  \right)\delta(z-x),
\end{equation}
\begin{equation}
\alpha=\left( \begin{array}{cc}
-\phi^{\prime\prime}-\dot {A_{1} }+A_{0}^{\prime}\\
{A_{0}}^{\prime \prime}-\dot{A_{1}}^{\prime}- {\phi}^{\prime} -ae^{2}A_{0}-ae\dot{\theta} \\
-\dot {A_{0}}^{\prime}+ae^{2}A_{1}+\dot{\phi}+ae\theta^{\prime}\\
-a\theta^{\prime\prime}+ae\dot{A_{0}}-aeA_{1}^{\prime}
\end{array}  \right).
\end{equation}
We see that Hessian matrix $w$ possesses the following null eigen
vector
\begin{equation}
\lambda^{1}(z) =\left( \begin{array}{rr} 0,1,0,0
\end{array}  \right) \delta(z-x),
\end{equation}
and the primary lagrangian constraint is obtained  here in the
same way by Multiplying $\lambda^{1}$ with (\ref{bu}) from the
left.
\begin{equation}
\gamma^{1}(x,t)=({A_{0}}^{\prime
\prime}-\dot{A_{1}}^{\prime}-ae^{2}A_{0}-
{\phi}^{\prime}-ae\dot{\theta})(x,t).
\end{equation}
In order to maintain consistency of the above primary constraint we add
 the time derivatives of $\gamma^{1}$  with $L_{ECM}$ and obtain $L_{1}(x,t)$ as follows.
\begin{equation}
L_{1}(x,t)=(L_{ESM}+\frac{\partial}{\partial t}\gamma^{1})(x,t).
\label{bu2}
\end{equation}
The above $L_{1}(x,t)$ contains $ w^{1}$ and $\alpha^{1}$ as usual
which are given by
\begin{equation}
w^{1}=\left( \begin{array}{rrrr}
1 & 0 & 0 &0\\
0 & 0 & 0&0\\
0&0&1&0\\
0&0&0&+a\\
0& 0&-\partial/ \partial x&-ae
\end{array}  \right)\delta(z-x),
\end{equation}
\begin{equation}
\alpha^{1}=\left( \begin{array}{cc}
-\phi^{\prime\prime}-\dot {A_{1} }+A_{0}^{\prime}\\
{A_{0}}^{\prime \prime}-\dot{A_{1}}^{\prime}- {\phi}^{\prime}
-ae^{2}A_{0}-ae\dot{\theta} \\ -\dot
{A_{0}}^{\prime}+ae^{2}A_{1}+\dot{\phi}+ae\theta^{\prime}
\\
-a\theta^{\prime\prime}+ae\dot{A_{0}}-aeA_{1}^{\prime}\\
\dot{A_{0}}^{\prime \prime}- \dot{\phi}^{\prime}
-ae^{2}\dot{A_{0}}
\end{array}  \right).
\end{equation}
The  matrix $w^{1}$ in this situation is found to give the
following null eigen vector.
\begin{equation}
\lambda^{2}(z) =\left( \begin{array}{rr} 0,0,\partial/ \partial
x,e,1,
\end{array}  \right)\delta(z-x).
\end{equation}
Multiplying the equation (\ref{bu2}) from left by $\lambda^{2} $,
we get
\begin{equation}
\gamma^{2}(x,t)=0. \label{bu3}
\end{equation}
The above vanishing condition establishes that multiplication of
$\lambda^{2}$ with $ L_{1}$ will not give  new  constraint.
Naturally,
 rank of equation for acceleration will not be increased.
 If we write equation (\ref{bu3}) in the form of equation (\ref{lambda}) as done in
 earlier different cases  it looks
\begin{equation}
\gamma^{2}(x,t)=(\frac {\partial}{\partial t} L_{A_{0}}+\frac
{\partial}{\partial x} L_{A_{1}}+eL_{\theta})(x,t). \label{bu4}
\end{equation}
One needs to Compare equation (\ref{bu4}) with (\ref{ph}) to
obtain $\phi_{si}$'s and that results the following.
\begin{equation}
\phi_{0,1}(z,x)=0,
\end{equation}
\begin{equation}
\phi_{1,2}(z,x)=\delta(z-x),
\end{equation}
\begin{equation}
\phi_{0,3}(z,x)=\frac{\partial}{\partial z}\delta(z-x),
\end{equation}
\begin{equation}
\phi_{0,4}(z,x)=e\delta(z-x).
\end{equation}
The gauge transformation for the field $\phi$, $A_{0}$, $A_{1}$
and $\theta$ formula that follows from the formula (\ref{GTF}) are
\begin{equation}
\delta \phi=0,  \label{bu5}
\end{equation}
\begin{equation}
\delta A_{0}=-\frac{\partial }{\partial t}f(x,t),
\label{bu6}
\end{equation}
\begin{equation}
\delta A_{1}=-\frac{\partial }{\partial x}f(x,t), \label{bu7}
\end{equation}
\begin{equation}
\delta \theta=ef(x,t).
\label{bu8}
\end{equation}
The variation of ${\cal L}_{ECM}$ under the above transformations
(\ref{bu5}), (\ref{bu6}), (\ref{bu7}) and (\ref{bu8}) is found out
to be
\begin{eqnarray}
\delta L(x,t)_{ESM}&=&\sum^{n}_{s=0}\frac{\partial^{s}}{\partial t^{s}}( \Phi_{si}L_{i})f(x,t)\nonumber\\
&=&(\frac {\partial}{\partial t} L_{A_{0}}+\frac
{\partial}{\partial x} L_{A_{1}}+eL_{\theta})f(x,t)=0.
\end{eqnarray}
It shows that in the extended phase space the lagrangian
(\ref{qa}) is invariant under the gauge transformation
(\ref{bu5}), (\ref{bu6}), (\ref{bu7}) and (\ref{bu8}). Therefore,
it is found that the formalism is capable of inferring about the
gauge symmetric property of this interacting system in the
extended phase space too. We will now proceed to consider the
another interacting system the so called Chiral Schwinger model
with the Faddevian anomaly in the following section.
\subsection{Chiral Schwinger Model with Fadeevian anomaly} Let us
consider the lagrangian of the so called Chiral Schwinger model
with Faddevian anomaly \cite{lag10, lag11} and apply the same
formalism to study its gauge symmetric property. The gauss law of
this theory shows a special type of non vanishing commutation
relation because of the presence of anomaly in the system. This is
commonly known as Faddeevian type of anomaly \cite{fad1, fad2}.
This model is interesting in different respect \cite{lag10,
lag11}. So study of this model with this formalism would certainly
be of interest.
\begin{eqnarray}
L_{CSM}&=&\int(-\frac{1}{4}F_{\mu\nu}F^{\mu\nu}+\frac{1}{2}\partial_{\mu}\phi
\partial^{\mu}\phi+e(g_{\mu\nu}-\epsilon_{\mu\nu})\partial^{\mu}\phi
A^{\nu}
 \nonumber\\&+&\frac{1}{2}e^{2}(A_{0}^{2}
-2A_{0}A_{1}-3 A_{1}^{2}))dx.
\end{eqnarray}
Explicitly the lagrangian can be written down as follows
\begin{eqnarray}
L_{CSM}=\int[&+&\frac{1}{2}(\dot{A_{1}}-
A_{0}^{\prime})^{2})+\frac{1}{2}(\dot{\phi^{2}}-\phi^{\prime^{2}})
 \nonumber\\&+&\frac{1}{2}e^{2}(A_{0}^{2}-2 A_{0}A_{1}-3 A_{1}^{2})
 +e(\dot{\phi}+\phi^{\prime})(A_{0}-A_{1})]dx. \label{ano}
\end{eqnarray}
The equations of motion for the field $\phi$, $ A_{0}$ and $
A_{1}$ are
\begin{equation}
L_{\phi}=(\ddot {\phi}-\phi^{\prime \prime})
+e(A_{0}^{\prime}-A_{1}^{\prime})+e(\dot{A_{0}}-\dot{A_{1}}),
\end{equation}
\begin{equation}
L_{A_{0}}={A_{0}}^{\prime \prime}-\dot{A_{1}}^{\prime}-e(
{\phi}^{\prime}+\dot{\phi})-e^{2}A_{0}+e^{2}A_{1},
\end{equation}
\begin{equation}
L_{A_{1}}=\ddot {A_{1}}-\dot
{A_{0}}^{\prime}+3e^{2}A_{1}+e({\phi}^{\prime}+\dot{\phi})+e^{2}A_{0}.
\end{equation}
For this lagrangian the Hessian matrix $w$ and $\alpha$ are
\begin{equation}
w=\left( \begin{array}{rrr}
1& 0 & 0 \\
0 & 0 & 0 \\
0 & 0 & 1\\
\end{array}  \right)\delta(z-x),
\end{equation}
\begin{equation}
\alpha=\left( \begin{array}{cc}
-\phi^{\prime\prime}+e(\dot{A_{0}}-\dot {A_{1} })+e(A_{0}^{\prime}-A_{1}^{\prime})\\
{A_{0}}^{\prime \prime}-\dot{A_{1}}^{\prime}-e(\dot{\phi}+ {\phi}^{\prime}) -e^{2}A_{0} +e^{2}A_{1}\\
 -\dot {A_{0}}^{\prime}+3e^{2}A_{1}+e(\dot{\phi}+\phi^{\prime})+e^{2}A_{0}\\
\end{array}  \right).
\end{equation}
A little algebra shows that the Hessian matrix $w$ bears the
following null eigen vector
\begin{equation}
\lambda^{1}(z) =\left( \begin{array}{rr} 0,1,0
\end{array}  \right)\delta(z-x).
\end{equation}
Here too multiplying  equation (\ref{ano}) from left by
$\lambda^{1} $ we get the primary constraint,
\begin{equation}
\gamma^{1}(x,t)=({A_{0}}^{\prime
\prime}-\dot{A_{1}}^{\prime}-e^{2}A_{0}+e^{2}A_{1}-e(\dot{\phi}+{\phi}^{\prime}))(x,t).
\end{equation}
The constraint has to be consistent with time. So we add time
derivatives of $\gamma^{1}$  with $L_{CSM}$, and it results
\begin{equation}
L_{1}(x,t)=(L_{CSM}+\frac{\partial}{\partial t}\gamma^{1})(x,t).
\label{ano3}
\end{equation}
In matrix  $w^{1}$ and $\alpha^{1}$ that occurs in equation (\ref{ano}) are found out as
\begin{equation}
w^{1}=\left( \begin{array}{rrr}
1 & 0 &0\\
0 & 0 & 0 \\
0& 0&1\\
-e& 0&-\partial/ \partial x
\end{array}  \right)\delta(z-x),
\end{equation}
\begin{equation}
\alpha^{1}=\left( \begin{array}{cc}
-\phi^{\prime\prime}+e(\dot {A_{0}}-\dot{A_{1}})+e(A_{0}^{\prime}-A_{1}^{\prime})\\
{A_{0}}^{\prime \prime}-\dot{A_{1}}^{\prime}- e(\dot{\phi}+{\phi}^{\prime}) -e^{2}A_{0}+e^{2}A_{1} \\ -\dot {A_{0}}^{\prime}+3e^{2}A_{1}+e^{2}A_{0}+e(\dot{\phi}+\phi^{\prime})\\
\dot {A_{0}}^{\prime\prime}-e^{2}\dot {A_{0}}-e^{2}\dot\phi^{\prime}+e^{2}\dot{A_{1}}\\
\end{array}  \right).
\end{equation}
Let us now calculate null vector $\lambda^{2}(z)$ which the matrix
$w^{1}$ is having.
\begin{equation}
\lambda^{2}(z) =\left( \begin{array}{rr} e,0,\partial/ \partial
x,1
\end{array}  \right)\delta(z-x).
\end{equation}
We get secondary lagrangian constraint multiplying equation
(\ref{ano3}) from left by $\lambda^{2}$:
\begin{equation}
\gamma^{2}(x,t)=2e^{2}(A_{1}^{\prime}+A_{0}^{\prime})(x,t).
\end{equation}
In order to maintain consistency again the time derivatives of
$\gamma^{2}$  is added with $ L_{1}$ which results
\begin{equation}
L_{2}(x,t)=(L_{1}+\frac{\partial}{\partial t}\gamma^{2})(x,t).
\label{last}
\end{equation}
In this situation the matrices $w^{2}$ and $\alpha^{2}$ for
$L_{2}(x,t)$ are found out as
\begin{equation}
w^{2}=\left( \begin{array}{rrr}
1 & 0 &0\\
0 & 0 & 0 \\
0& 0&1\\
-e& 0&-\partial/ \partial x\\
0 &0&0
\end{array}  \right)\delta(z-x),
\end{equation}
\begin{equation}
\alpha^{2}=\left( \begin{array}{cc}
-\phi^{\prime\prime}+e(\dot {A_{0}}-\dot{A_{1}})+e(A_{0}^{\prime}-A_{1}^{\prime})\\
{A_{0}}^{\prime \prime}-\dot{A_{1}}^{\prime}- e(\dot{\phi}+{\phi}^{\prime}) -e^{2}A_{0}+e^{2}A_{1} \\ -\dot {A_{0}}^{\prime}+3e^{2}A_{1}+e^{2}A_{0}+e(\dot{\phi}+\phi^{\prime})\\
\dot {A_{0}}^{\prime\prime}-e^{2}\dot {A_{0}}-e^{2}\dot\phi^{\prime}+e^{2}\dot{A_{1}}\\
2e^{2}(\dot{A_{1}^{\prime}}+\dot{A_{0}^{\prime}})
\end{array}  \right).
\end{equation}
A little algebra shows that $w^{2}$ also has a new null eigen vector
\begin{equation}
\lambda^{3}(z) =\left( \begin{array}{rr} e,0,\partial/ \partial
x,1,0
\end{array}  \right) \delta(z-x).
\end{equation}
If we
 multiplying the equation (\ref{last}) from left by
$\lambda^{3}$ and get
\begin{equation}
\gamma^{3}(x,t)=2e^{2}(A_{1}^{\prime}+A_{0}^{\prime})(x,t)=\gamma^{2}(x,t).
\label{ano8}
\end{equation}
The mapping of $\gamma^{3}(x,t)$ onto $\gamma^{2}(x,t)$ indicates
that there is no other constraint. Thus the process is terminated.
As it is done in this previous cases $\gamma^{3}$ here too is
expressed in the form of equation (\ref{lambda}).
\begin{equation}
\gamma^{3}(x,t)=(\frac {\partial}{\partial t} L_{A_{0}}+\frac
{\partial}{\partial x} L_{A_{1}}+eL_{\phi})(x,t).
\end{equation}
Comparing $\gamma^{3}(x,t)$ with equation (\ref{ph}) we get non
vanishing $\phi_{si}$'s which will be useful to calculate gauge
transformations of the fields describing the system.
\begin{equation}
\phi_{0,1}(z,x)=e\delta(z-x),
\end{equation}
\begin{equation}
\phi_{1,2}(z,x)=\delta(z-x),
\end{equation}
\begin{equation}
\phi_{0,3}(z,x)=\frac{\partial}{\partial x}\delta(z-x).
\end{equation}
We are now in a state to find the gauge transformation of the
field describing the system by using equation (\ref{GTF}).
\begin{equation}
\delta \phi=ef(x,t),\label{TT0}
\end{equation}
\begin{equation}
\delta A_{0}=-\frac{\partial }{\partial t}f(x,t), \label{TT1}
\end{equation}
\begin{equation}
\delta A_{1}=-\frac{\partial }{\partial x}f(x,t).\label{TT2}
\end{equation}
The variation of $L_{CSM}$ under the transformations (\ref{TT0}),
(\ref{TT1}) and (\ref{TT1} are found out to be
\begin{eqnarray}
\delta{\cal L}_{CSM}
&=&-(\sum^{n}_{s=0}\frac{\partial^{s}}{\partial t^{s}}(
\Phi_{si}L_{i})f(x,t)\nonumber\\&=&-(eL_{\phi}
+\frac {\partial}{\partial t} L_{A_{0}}+\frac {\partial}{\partial x} L_{A_{1}})f(x,t)\nonumber\\
&=&-2e^{2}(A_{1}^{\prime}+{A_{0}^{\prime}})f(x,t). \label{NC}
\end{eqnarray}
So it is found that the lagrangian (\ref{ano}) is not invariant
under the above transformations. The formalism here too gives the
expected result because it is an anomalous model with Faddeevian
type of anomaly and the appearance of  gauge non invariance for
this model is obvious.
\subsection{Chiral Schwinger Model with Fadeevian anomaly made Gauge invariant in the extended phase space}
The lagrangian of Chiral Schwinger model with Fadeevian anomaly is
found gauge non invariant under the transformation generated in
the previous section by Shirzad's formalism. So we add some terms
with the previous lagrangian (\ref{ano}) to bring back its
symmetry and apply the prescription to verify whether we get the
expected result in the extended phase space like the previous
case. Lagrangian density with appropriate terms involving the
auxiliary field $\theta$ that helps to make equation (\ref{NC})
zero reads
\begin{eqnarray}
{\cal L}_{ECSM}=&+&\frac{1}{2}(\dot{A_{1}}-
A_{0}^{\prime})^{2})+\frac{1}{2}(\dot{\phi^{2}}-\phi^{\prime^{2}})+\frac{1}{2}e^{2}(A_{0}^{2}-2
A_{0}A_{1}-3 A_{1}^{2})
\nonumber\\
     &+&e(\dot{\phi}+\phi^{\prime})(A_{0}-A_{1})+\frac{1}{2}(\dot{\theta^{2}}-2\dot{\theta}\theta^{\prime}
-3\theta^{\prime^{2}})\nonumber\\
&-&\frac{1}{2}(\dot{\theta}^{2}-\theta^{\prime^{2}})-e(A_{0}\dot{\theta}-A_{0}\theta^{\prime}
-A_{1}\dot{\theta}-3A_{1}\theta^{\prime}) \nonumber\\
&+&e(A_{0}\theta^{\prime}-A_{1}\dot{\theta})+e(-A_{1}\theta^{\prime}+A_{0}\dot{\theta}).\label{end}
\end{eqnarray}
The equations of motion corresponding to the field $\phi$, $
A_{0}$, $ A_{1}$ and $\theta$ are
\begin{equation}
L_{\phi}=(\ddot {\phi}-\phi^{\prime \prime})
+e(A_{0}^{\prime}-A_{1}^{\prime})+e(\dot{A_{0}}-\dot{A_{1}}),
\end{equation}
\begin{equation}
L_{A_{0}}={A_{0}}^{\prime \prime}-\dot{A_{1}}^{\prime}-e(
{\phi}^{\prime}+\dot{\phi})-e^{2}A_{0}+e^{2}A_{1}-2e\theta^{\prime},
\end{equation}
\begin{equation}
L_{A_{1}}=\ddot {A_{1}}-\dot
{A_{0}}^{\prime}+3e^{2}A_{1}+e({\phi}^{\prime}+\dot{\phi})+e^{2}A_{0}-2e\theta^{\prime},
\end{equation}
\begin{equation}
 L_{\theta}=-2\theta^{\prime\prime}-2\dot{\theta}^{\prime}+2eA_{0}^{\prime}
+2eA_{1}^{\prime}.
\end{equation}
The matrices $w$ and $\alpha$ in this situation are
\begin{equation}
w=\left( \begin{array}{rrrr}
1& 0 & 0&0 \\
0 & 0 & 0&0 \\
0 & 0 & 1&0\\
0& 0& 0& 0\\
\end{array}  \right)\delta(z-x),
\end{equation}
\begin{equation}
\alpha=\left( \begin{array}{cc}
-\phi^{\prime\prime}+e(\dot{A_{0}}-\dot {A_{1} })+e(A_{0}^{\prime}-A_{1}^{\prime})\\
{A_{0}}^{\prime \prime}-\dot{A_{1}}^{\prime}-e(\dot{\phi}+ {\phi}^{\prime}) -e^{2}A_{0}
+e^{2}A_{1}-2e\theta^{\prime}\\ -\dot {A_{0}}^{\prime}+e(\dot{\phi}+\phi^{\prime})+e^{2}A_{0}
-2e\theta^{\prime}+3e^{2}A_{1}\\
-2\theta^{\prime\prime}-2\dot{\theta}^{\prime}+2e(A_{0}^{\prime}+A_{1}^{\prime})
\end{array}  \right).
\end{equation}
It is found that there exists a null vector within the the Hessian matrix $w$ which is given by
\begin{equation}
\lambda^{1}(z) =\left( \begin{array}{rr} 0,1,0,0
\end{array}  \right)\delta(z-x).
\end{equation}
Multiplying the equation (\ref{end}) with $\lambda^{1}$ from the
left we obtain the following primary constraint
\begin{equation}
\gamma^{1}(x,t)=({A_{0}}^{\prime
\prime}-\dot{A_{1}}^{\prime}-e^{2}A_{0}+e^{2}A_{1}-e(\dot{\phi}+
{\phi}^{\prime})-2e\theta^{\prime})(x,t).
\end{equation}
When the time derivative of $\gamma^{1}$ is added with $L_{ESM}$
it gives $ L_{1}$ which is the requirement for the primary
constraint to be consistent with time.
\begin{equation}
L_{1}(x,t)=(L_{ECSM}+\frac{\partial}{\partial t}\gamma^{1})(x,t).
\label{end12}
\end{equation}
It is now needed to find out $w^{1}$ and $\alpha^{1}$ contained in
$L_{1}(x,t)$ for this situation.
\begin{equation}
w^{1}=\left( \begin{array}{rrrr}
1 & 0 &0&0\\
0 & 0 & 0&0 \\
0& 0&1&0\\
-e& 0&-\partial/ \partial x&0
\end{array}  \right)\delta(z-x),
\end{equation}
\begin{equation}
\alpha^{1}=\left( \begin{array}{cc}
-\phi^{\prime\prime}+e(\dot{A_{0}}-\dot {A_{1} })+e(A_{0}^{\prime}-A_{1}^{\prime})\\
{A_{0}}^{\prime \prime}-\dot{A_{1}}^{\prime}-e(\dot{\phi}+ {\phi}^{\prime}) -e^{2}A_{0}
+e^{2}A_{1}-2e\theta^{\prime}\\ -\dot {A_{0}}^{\prime}+e(\dot{\phi}+\phi^{\prime})
+e^{2}A_{0}-2e\theta^{\prime}+3e^{2}A_{1}\\
-2\theta^{\prime\prime}-2\dot{\theta}^{\prime}+2e(A_{0}^{\prime}+A_{1}^{\prime})\\
\dot{A_{0}}^{\prime \prime}-e\dot {\phi}^{\prime}
-e^{2}\dot{A_{0}} +e^{2}\dot{A_{1}}-2e\dot{\theta}^{\prime}
\end{array}  \right).
\end{equation}
Our next step is to find out whether the matrix do have any null
eigen vector and it is seen that matrix $ w^{1}$ has the following
null eigen vector.
\begin{equation}
\lambda^{2}(z) =\left( \begin{array}{rrr} e,0,\partial/ \partial
x,-e,1
\end{array}  \right) \delta(z-x).
\end{equation}
Now we multiply the equation (\ref{end12}) from left by
$\lambda^{2}$ to find $\gamma^{2}$ which turns out to be zero
here.
\begin{equation}
\gamma^{2}(x,t)=\lambda^{2}\alpha_{2}(x,t)=0.
\end{equation}
So it is not possible to increase the the rank of equation for
accelerations. Since $\lambda^{2}$ does not give rise to any new
constraint one needs to express  $\gamma^{2}$ in the  form of equation
(\ref{lambda}).
\begin{equation}
\gamma^{2}(x,t)=(\frac {\partial}{\partial t} L_{A_{0}}+\frac
{\partial}{\partial x}
L_{A_{1}}+eL_{\phi}-eL_{\theta})(x,t).\label{end13}
\end{equation}
Equating (\ref{end13}) with equation (\ref{ph}) we find that the
non vanishing $\phi_{si}$'s in this situation are
\begin{equation}
\phi_{0,1}(z,x)=e\delta(z-x),
\end{equation}
\begin{equation}
\phi_{1,2}(z,x)=\delta(z-x),
\end{equation}
\begin{equation}
\phi_{0,3}(z,x)=\frac{\partial}{\partial z}\delta(z-x),
\end{equation}
\begin{equation}
\phi_{0,4}(z,x)=-e\delta(z-x).
\end{equation}
It is now straightforward to compute the gauge transformations
(\ref{GTF})
 for the field $\phi$, $A_{0}$  $A_{1}$ and $\theta$.
\begin{equation}
\delta \phi=ef(x,t), \label{end14}
\end{equation}
\begin{equation}
\delta A_{0}=-\frac{\partial }{\partial t}f(x,t),\label{end15}
\end{equation}
\begin{equation}
\delta A_{1}=-\frac{\partial }{\partial x}f(x,t),\label{end16}
\end{equation}
\begin{equation}
\delta \theta=-ef(x,t).\label{end17}
\end{equation}
A little algebra shows that the variation of ${\cal L}_{ECSM}$
under the above gauge transformations is
\begin{equation}
\delta {\cal L}_{ECSM}=(\frac {\partial}{\partial t}
L_{A_{0}}+\frac {\partial}{\partial x}
L_{A_{1}}+eL_{\phi}-eL_{\theta})f(x,t) =0.
\end{equation}
It  shows that the lagrangian (\ref{end}) is invariant under the
gauge transformation equations (\ref{end14}), (\ref{end15}),
(\ref{end16}) and (\ref{end17}). Therefore, we again observe the
real ability of this formalism for testing the gauge symmetric
property in the extended phase space of the so called chiral
Schwinger model with Faddeevian anomaly.
\section{Conclusion}
An instrument for testing gauge symmetry as well as generating
gauge transformation of a theory through lagrangian formulation
developed by Shirzad in \cite{lag23} has been applied on different
interacting and non interacting field theoretical model. Some of
the model had gauge symmetry to start with and in some model it
was lacking. The formalism is found instrumental to study the
gauge symmetric property for all the cases whatever subtleties are
involved in these. Using Shirzad's formalism, we have successfully
tested whether a given model does posses gauge symmetry or not.
When a model is found gauge non-invariant it is made gauge
invariant by adding some auxiliary fields with the lagrangian of
that model and investigation is carried over using Shirzad's
prescription to test whether gauge symmetric gets restored in it.
The process of adding auxiliary fields though extends the phase
space the physical content of the theory remains unaltered because
the fields required for the extension keep themselves allocated in
the unphysical sector of the theory. More importantly, it has been
possible to generate gauge transformation generator in the
extended phase space too. So it is found that the formalism is not
only useful in the usual phase space of the theory but also it is
equally powerful in the extended phase space. In this context, we
should mention that  in \cite{lag23}, Shirzad kept himself
confined within the usual phase space of the theory.  One
important aspect of this formalism which we have noticed here is
that one can have a guess about the Weiss-Zumino term needed to
bring back the symmetry of a gauge non invariant theory. In every
cases of our studies we have noticed that the terms involving
auxiliary field needed for making the variation of a particular
gauge non invariant lagrangian zero under the respective
transformations generated through Shirzad's formalism leads to the
Weiss- Zumino term for the respective theory. But it is fair to
admit that the formalism is still lacking the mechanism to make a
theory gauge invariant in a straightforward manner, i.e., the
automatic generation of Wess-Zumino term as it has been found to
be generated during the BRST invariant reformulation through
Batalin-Fradkin-Vilkovisky formalism \cite{lag15,lag16,lag17}.
However the formalism has enough room for improvement towards this
end. More serious and intense investigation is needed in that
direction.

\end{document}